\documentclass{article}
\usepackage{graphicx}
\usepackage{amsmath}

\begin{document}

\title{Topological phase for entangled two-qubit states and the representation of
the $SO(3)$ group}
\author{F. De\ Zela \\
%EndAName
Departamento de Ciencias, Secci\'{o}n F\'{i}sica \\
Pontificia Universidad Cat\'{o}lica del Per\'{u}, Ap.1761, Lima, Per\'{u}.}
\maketitle

\begin{abstract}
We discuss the representation of the $SO(3)$ group by two-qubit maximally
entangled states (MES). We analyze the correspondence between $SO(3)$ and
the set of two-qubit MES which are experimentally realizable. As a result,
we offer a new interpretation of some recently proposed experiments based on
MES. Employing the tools of quantum optics we treat in terms of two-qubit
MES some classical experiments in neutron interferometry, which showed the $%
\pi $-phase accrued by a spin-$1/2$ particle precessing in a magnetic field.
By so doing, we can analyze the extent to which the recently proposed
experiments - and future ones of the same sort - would involve essentially
new physical aspects as compared with those performed in the past. We argue
that the proposed experiments do extend the possibilities for displaying the
double connectedness of $SO(3)$, although for that to be the case it results
necessary to map elements of $SU(2)$ onto physical operations acting on
two-level systems.

\end{abstract}

\section{Introduction}

Geometric phases and entangled states have been two of the most addressed
issues during the last few years. The widespread interest on these topics
embraces both theoretical and applied aspects. Since the appearance of
Berry's seminal paper \cite{berry}, different geometric phases have been
theoretically introduced and experimentally tested. Due to some subtleties
that are inherent to the above mentioned topics, the introduction of a new
phase and/or its experimental demonstration eventually causes some
controversy and it may take some time before an issue becomes settled. This
usually requires seeing one and the same topic from different perspectives.
It is in this vein that the present article deals with a topological phase
recently proposed by Milman and Mosseri \cite{milman}. These authors
discussed a quantum optics interference experiment designed to demonstrate a
topological phase shift, which should stem from the subtleties of the $SO(3)$
rotation group topology. Milman and Mosseri claim that by using two-qubit,
maximally entangled states (MES) the double connectedness of the $SO(3)$
rotation group can be displayed. Their proposal is based on a one-to-one map
between $SO(3)$ and the set of two-qubit MES. These states can be currently
produced as, e.g., polarization-entangled photon pairs by using a down
converter \cite{galvez}. In order to measure the topological phase, one of
the photons from the entangled pair is sent through a Mach-Zehnder
interferometer, at whose output it is detected in coincidence with the other
photon from the entangled pair, which serves as the reference mode. In one
of the arms of the Mach-Zehnder interferometer a variable dephasing $\phi $
is introduced. The goal is to bring into evidence that there are two classes
of trajectories in $SO(3)$, as followed by the MES when they are subjected
to the transformations due to different devices that constitute the optical
system (beam-splitters, wave plates, mirrors, etc.). As a result, the
counting rate for coincidence detections at two detectors, $D_{1}$ and $%
D_{2} $, put at the output ports of the interferometer, is given by $%
P=(1/2)\left| \left( -1\right) ^{n}-\cos \phi \right| $. Here, $n$ counts
the number of times that the trajectory breaks on the surface of the $SO(3)$
sphere, on which such trajectories can be represented. In a sequel to Ref.
\cite{milman}, LiMing, Tang and Liao \cite{liming} refine the proposal of
Milman and Mosseri, by showing how to realize a larger family of closed
trajectories, in which the sudden jumps of the original proposal are
replaced by a continuously changing path. Moreover, $n$ can have odd and
even values beyond the two only cases, $n=0$ and $n=1$, that Milman and
Mosseri had considered. To the best of our knowledge, the proposed
experiments have not yet been implemented.

One goal of the present work is to discuss the extent to which the proposal
of Milman and Mosseri, or that of LiMing, Tang and Liao, would imply the
appearance of a new phase. Around three decades ago a series of experiments
in neutron interferometry were performed \cite{werner}, whose aim was to
bring into evidence the $\pi $-phase that arises when a spinor is subjected
to a $2\pi $ rotation. Such a $\pi $-phase illustrates one of the most
salient features characterizing the interplay between the groups $SO(3)$ and
$SU(2)$. The advent of quantum optics allowed not only to extend our
capability to perform new experiments on this line, but helped also to
interpret the old experiments in the framework of entangled states. In this
work we present an alternative interpretation of experiments similar to
those proposed by Milman and Mosseri and LiMing, Tang and Liao. To this end,
we re-analyze the neutron experiments by employing the tools of quantum
optics, thereby recovering the known results but at the same time extending
their scope. Though we refer specifically to neutron experiments, our
results are valid for experiments that can be performed in quantum optics.
Only minor changes are needed to obtain the corresponding predictions when
one works with photons instead of dealing with fermions.

This paper is organized as follows. We first present the evolution of
maximally entangled states in connection with the $SU(2)$ group. We briefly
discuss those features concerning the double connectedness of $SO(3)$, which
are relevant for our case. A more detailed discussion, including some
technical details of $SO(3)$ has been included in the Appendix. This
Appendix should serve to make the present paper more self contained. In the
next section we discuss neutron interferometry by means of a full
quantum-mechanical treatment. This gives us the opportunity to stress the
essential role played by entangled states in this context. We analyze in
detail how the $\pi $-phase arises and compare our approach with another,
newly proposed one, that makes use of the geometrical phase for mixed
states. We then discuss the basic experimental arrangement - a Mach-Zehnder
interferometer - necessary to bring into evidence those features of the $%
SO(3)$ group that have been on the focus of our interest. We summarize our
results at the end of the paper and make some conclusions regarding the
novelty of the recent proposals.

\section{The topological phase}

Two-qubit MES are given by

\begin{equation}
\left| \alpha ,\beta \right\rangle =\frac{1}{\sqrt{2}}\left( \alpha \left|
0,0\right\rangle +\beta \left| 0,1\right\rangle -\beta ^{\ast }\left|
1,0\right\rangle +\alpha ^{\ast }\left| 1,1\right\rangle \right) ,  \label{1}
\end{equation}
with $\left| \alpha \right| ^{2}+\left| \beta \right| ^{2}=1$. Following
Milman and Mosseri we define the ``Hilbert space of all MES'' as $\Omega
_{MES}=\left\{ \left( \alpha ,\beta \right) \in C^{2}/\left| \alpha \right|
^{2}+\left| \beta \right| ^{2}=1\text{ and }\left( \alpha ,\beta \right)
\sim \left( -\alpha ,-\beta \right) \right\} =S^{3}/Z_{2}=SO(3)$. First of
all, let us note that what we obtain as a consequence of the identification $%
\left( \alpha ,\beta \right) \sim \left( -\alpha ,-\beta \right) $ is the
well known two-to-one homomorphism between $SU(2)$ $(=S^{3})$ and $SO(3)$ $%
(=S^{3}/Z_{2})$. Our main purpose here is to elucidate whether the
mathematical construction leading to $\Omega _{MES}$ has a corresponding
physical realization or not. The key point is the identification $\left(
\alpha ,\beta \right) \sim \left( -\alpha ,-\beta \right) $. It certainly
reflects the physical indistinguishability between the states $\left| \alpha
,\beta \right\rangle $ and $\left| -\alpha ,-\beta \right\rangle =e^{i\pi
}\left| \alpha ,\beta \right\rangle $. We note that, in general, $\left|
\alpha ,\beta \right\rangle \neq \left| e^{i\psi }\alpha ,e^{i\psi }\beta
\right\rangle $, so that one could expect that the production of MES in the
laboratory automatically leads to a physical realization of $\Omega _{MES}$.
However, considering the experimental production of MES, we realize that
what we can produce in the laboratory are states whose global phase remains
undetermined. This means that not only $(\alpha ,\beta )$ and $(-\alpha
,-\beta )$ are experimentally indistinguishable, but also $\left| \alpha
,\beta \right\rangle $ and any other state of the form $e^{i\psi }\left|
\alpha ,\beta \right\rangle $. Thus, instead of $\Omega _{MES}$, and because
the set whose elements are the factors $e^{i\psi }$ is nothing but $U(1)$ $%
(=S^{1})$, what we actually produce in the laboratory seems to be rather
akin to a realization of the Hopf-fibration \cite{mosseri} $S^{3}\overset{%
S^{1}}{\longrightarrow }S^{2}$, by which the $SU(2)$ manifold $(S^{3})$ is
locally decomposed in the form of products of $S^{2}$-elements $\left(
\alpha ,\beta \right) $ by elements $e^{i\psi }$ of the fiber $S^{1}$. The
precise way in which such a realization can be established is however not a
matter of our concern here. For our purposes, it should suffice to point out
that experimentally realizable MES are defined up to a global phase and can
be put in one-to-one correspondence with the $SU(2)$ elements.

Indeed, MES given by Eq.(\ref{1}) with $\left| \alpha \right| ^2+\left|
\beta \right| ^2=1$ can also be represented in matrix form as

\begin{equation}
M_{\left( \alpha ,\beta \right) }=\left(
\begin{tabular}{ll}
$\alpha $ & $\beta $ \\
$-\beta ^{*}$ & $\alpha ^{*}$%
\end{tabular}
\right) ,  \label{2}
\end{equation}
leaving aside the normalization factor $1/\sqrt{2}$. $M_{\left( \alpha
,\beta \right) }$ has thus the form of a general $SU(2)$-matrix. It is well
known that any Hamiltonian of a two-level system can be written in a form
that corresponds to the interaction of a spin-$1/2$ particle with a magnetic
field: $H=-\mathbf{\mu }\cdot \mathbf{B}\equiv \hbar \omega \mathbf{\sigma }%
\cdot \mathbf{n}/2$. If such a Hamiltonian drives the evolution of one of
the two qubits in Eq.(\ref{1}), the result is an evolved, two-qubit state $%
\left| \alpha (t),\beta (t)\right\rangle $ that remains a MES. Let us first
restrict ourselves to the case of a field having a constant direction $%
\mathbf{n}$. When $H$ acts on the first qubit, the evolved state $\left|
\alpha (t),\beta (t)\right\rangle $ can be obtained from the initial state $%
\left| \alpha ,\beta \right\rangle \equiv \left| \alpha (0),\beta
(0)\right\rangle , $ as

\begin{equation}
\left(
\begin{tabular}{ll}
$\alpha (t)$ & $\beta (t)$ \\
$-\beta ^{\ast }(t)$ & $\alpha ^{\ast }(t)$%
\end{tabular}
\right) =\left(
\begin{tabular}{ll}
$A(t)$ & $B(t)$ \\
$-B^{\ast }(t)$ & $A^{\ast }(t)$%
\end{tabular}
\right) \left(
\begin{tabular}{ll}
$\alpha $ & $\beta $ \\
$-\beta ^{\ast }$ & $\alpha ^{\ast }$%
\end{tabular}
\right) \equiv M_{U}M_{\left( \alpha ,\beta \right) },  \label{3}
\end{equation}
whereby the first matrix on the right-hand side corresponds to the evolution
operator $U=\exp \left( -i\omega t\left( \mathbf{n}\cdot \mathbf{\sigma }%
\right) /2\right) =I\cos \left( \omega t/2\right) -i\mathbf{n}\cdot \mathbf{%
\sigma }\sin \left( \omega t/2\right) $ acting on the first qubit. One finds
that $A(t)$ and $B(t)$ are given by just the same elements of $U$:

\bigskip
\begin{eqnarray}
A(t) &=&\cos \left( \omega t/2\right) -in_{z}\sin \left( \omega t/2\right)
\label{4} \\
B(t) &=&-i\left( n_{x}-in_{y}\right) \sin \left( \omega t/2\right) .
\label{5}
\end{eqnarray}
\

A similar expression can be obtained for the case when it is the second
qubit the one which is subjected to the action of the Hamiltonian. In such a
case, $\left| \alpha (t),\beta (t)\right\rangle $ is given by a relation
similar to Eq.(\ref{3}), but involving $U^T$, the transpose of $U$:

\begin{equation}
\left(
\begin{tabular}{ll}
$\alpha (t)$ & $\beta (t)$ \\
$-\beta ^{\ast }(t)$ & $\alpha ^{\ast }(t)$%
\end{tabular}
\right) =\left(
\begin{tabular}{ll}
$\alpha $ & $\beta $ \\
$-\beta ^{\ast }$ & $\alpha ^{\ast }$%
\end{tabular}
\right) \left(
\begin{tabular}{ll}
$A(t)$ & $-B^{\ast }(t)$ \\
$B(t)$ & $A^{\ast }(t)$%
\end{tabular}
\right) =M_{\left( \alpha ,\beta \right) }(M_{U})^{T}  \label{6}
\end{equation}

As we see, the two evolutions bring an initial $SU(2)$-element into another
one, so that maximal entanglement is preserved. This is so because all the
involved matrices, $M_{\left( \alpha ,\beta \right) }$, $M_{U}$, and $%
(M_{U})^{T}$, are elements of $SU(2)$ and by multiplying two of them what we
obtain is just another element of the same group. Each matrix $M_{\left(
\alpha ,\beta \right) }$ can be seen as the result of applying a rotation $%
M_{U}$ with $A=\alpha $ and $B=\beta $ to the identity matrix $M_{\left(
\alpha =1,\beta =0\right) }$, which corresponds to the state $\left|
1,0\right\rangle $. In other words, $M_{\left( \alpha ,\beta \right) }$ can
be identified with the rotation $M_{U(\alpha ,\beta )}$. This way, two-qubit
MES span just another representation-space for rotations. Using it we obtain
a representation which is isomorphic to the usual, one-qubit, $SU(2)$ spinor
representation. For closed trajectories a $\pi $-phase might arise: e.g.,
when $\omega t=2\pi $, so that $\left| \alpha (t),\beta (t)\right\rangle
=-\left| \alpha ,\beta \right\rangle $ . This phase can only be \cite
{aharonov}, of course, a relative phase between two states (e.g., light
beams), whereby one of the states is taken as a reference state. By making
the two beams interfere we can bring the $\pi $-phase into evidence. This
was the aim of the classical experiments in neutron interferometry which
were developed in the past \cite{werner,rauch}. In the following Section we
shall discuss these experiments in the framework of maximally entangled
states, so as to bring to the fore those features that the classical
experiments have in common with the newly proposed ones. However, before
going into the analysis of the neutron experiments, it seems worthwhile to
briefly discuss those general matters related to the $SO(3)$ double
connectedness that have prompted some recent proposals, like the ones
referred to above \cite{milman,liming}. This will also set the stage for the
discussion that follows. A more detailed treatment of the $SO(3)$ double
connectedness and related issues has been included in the Appendix.

The double connectedness of $SO(3)$ refers to the fact that not every closed
curve in the $SO(3)$ parameter space can be continuously deformed to a
single point at the origin, which is the point representing the identity
transformation. In other words, not every closed curve in $SO(3)$ is
homotopically equivalent to a point. The curves which are homotopic to the
identity belong to one homotopy class. The curves which are not homotopic to
the identity belong to a second homotopy class. It happens that $SO(3)$ has
only two homotopy classes. Now, if we take an object and apply to it a
sequence of $SO(3)$ transformations which begin and end at the identity, we
cannot tell - by just observing the object - which homotopy class the
sequence belongs to. In order to identify the homotopy class we need to map
the sequence of transformations into elements of $SU(2)$, the simply
connected group which shares with $SO(3)$ the same Lie algebra. For each
element of $SO(3)$ there are two elements of $SU(2)$ that can be mapped on
it. Such a map allows us to disclose the double connectedness of $SO(3)$.
These topics are discussed in more detail in the Appendix. Here, it should
be enough to stress that in order to disclose the double connectedness of $%
SO(3)$ we do need an object on which to apply the corresponding $SU(2)$
transformations. Of course, such an object is nothing but a spin-$1/2$
particle or any other two-level system. We cannot remain working with $SO(3)$
transformations or with objects on which these transformations apply, if we
aim at disclosing the subtleties of the $SO(3)$ topology. If that would be
possible, a rigid body would suffice. Thus, the essential point concerning
MES is not that they could be put in one-to-one correspondence with a set of
$3$-dimensional rotation matrices - which is true for rigid bodies -, but
that MES entail two-level systems on which we can apply $SU(2)$
transformations. In order to disclose the subtleties of the $SO(3)$ topology
we need at least three things: spinors, entanglement, and the possibility of
interfering a ``rotated'' with a ``non-rotated'' spinor.

Milman and Mosseri \cite{milman} considered two sequences of
transformations. Both sequences start by being applied to the identity $%
(\alpha ,\beta )=(1,0)$ and consist of four rotations, whereby the first two
of these rotations act on the first qubit, and the last ones on the second
qubit. However, one sequence belongs to one homotopy class, bringing the
point $(1,0)$ back to itself, whereas the other sequence has $(-1,0)$ as its
endpoint and belongs to the other homotopy class. The idea is to consider
sequences of rotations for which the rotation axis is not kept constant, as
it was the case in the neutron experiments referred to above. According to
Milman and Mosseri, by keeping the rotation axis constant the $\pi $ phase
detected in those experiments is not necessarily related to the ``subtle
nature of $SO(3)$'', but to ``a property shared by $SO(3)$ and its subgroup $%
SO(2)$''. In order to apply those results which are valid only for a
constant rotation axis, the sequence of rotations considered in \cite{milman}
does not entail a continuously varying rotation axis, but it is split into
four rotations. Each one of these rotations is performed about a fixed
direction, which is changed from one rotation to the next. LiMing \emph{et al%
}. \cite{liming} consider instead a continuously changing rotation axis $%
\mathbf{n}(t)=(\sin \theta \cos \omega t,\sin \theta \sin \omega t,\cos
\theta )$. This particular case corresponds to a spin driven by a magnetic
field which is precessing around the $z$-axis. The corresponding
Schr\"{o}dinger equation can be solved exactly in this case as well \cite
{rabi,bohm}.

Now, for our present purposes, it suffices to note that the evolution of a
spinor driven by a general Hamiltonian $H(t)=\hbar \omega \mathbf{\sigma }%
\cdot \mathbf{n}(t)/2$ is given by $\left| \psi (t)\right\rangle =U(t)\left|
\psi (0)\right\rangle $, with $U(t)\in SU(2)$. This holds true for a general
$\mathbf{n}(t)$, even though an expression for $U(t)$ is not easily obtained
in closed form but in the two aforementioned cases. Hence, Eqs.(\ref{3}) and
(\ref{6}) remain valid for a variable $\mathbf{n}(t)$, in general. These
equations enable us to calculate - at least formally - the evolved MES as

\begin{equation}
\left| \alpha (t),\beta (t)\right\rangle =\frac{1}{\sqrt{2}}\left( \alpha
(t)\left| 0,0\right\rangle +\beta (t)\left| 0,1\right\rangle -\beta ^{\ast
}(t)\left| 1,0\right\rangle +\alpha ^{\ast }(t)\left| 1,1\right\rangle
\right) .
\end{equation}

Milman and Mosseri point out one aspect of the sequences of rotations they
consider. Namely, that for both sequences, irrespective of the homotopy
class to which they belong, their dynamical phase vanishes. As for the
Pancharatnam phase, it also vanishes, except in those cases where the
evolved state ``crosses the space orthogonal to the initial state, where it
abruptly changes by $\pi $'' \cite{milman}. This feature is interpreted as
giving further support to the contention that these $\pi $-phases have a
purely topological origin.

Taking the time derivative of Eq.(7) it is easy to see that not only for the
particular sequences considered by Milman and Mosseri, or for those proposed
by LiMing \emph{et. al}., but quite generally, the states $\left| \alpha
(t),\beta (t)\right\rangle $ evolve in time satisfying

\begin{equation}
\left\langle \alpha (t),\beta (t)\right| \frac{d}{dt}\left| \alpha (t),\beta
(t)\right\rangle =\frac{1}{2}\frac{d}{dt}\left( \left| \alpha (t)\right|
^{2}+\left| \beta (t)\right| ^{2}\right) =0,
\end{equation}
which is the condition for parallel transport \cite{anandan}. Furthermore,

\begin{equation}
\left\langle \alpha (0),\beta (0)|\alpha (t),\beta (t)\right\rangle
=Re\left( \alpha ^{\ast }(0)\alpha (t)+\beta ^{\ast }(0)\beta (t)\right) .
\end{equation}

Hence, recalling the definition of the geometric phase,

\begin{equation}
\Phi _g=\arg \left\langle \alpha (0),\beta (0)|\alpha (\tau ),\beta (\tau
)\right\rangle +i\int_0^\tau \left\langle \alpha (t),\beta (t)\right| \frac
d{dt}\left| \alpha (t),\beta (t)\right\rangle dt,
\end{equation}
we see that, because the dynamical phase $-i\int_0^\tau \left\langle \alpha
(t),\beta (t)\right| \frac d{dt}\left| \alpha (t),\beta (t)\right\rangle
dt=0 $ in our case, the geometric phase $\Phi _g$ reduces to the
Pancharatnam - or total - phase; i.e., $\Phi _g=\arg \left\langle \alpha
(0),\beta (0)|\alpha (\tau ),\beta (\tau )\right\rangle $. Whenever $\left|
\alpha (\tau ),\beta (\tau )\right\rangle =-\left| \alpha (0),\beta
(0)\right\rangle $, the geometric phase $\Phi _g=\pi $. So we see that each
time the point $\left( \alpha (t),\beta (t)\right) $ is brought to the
surface of the solid ball of radius $\pi $ that constitutes the parameter
space of $SO(3)$ (see the Appendix), it accumulates a $\pi $ phase. However,
we can hardly say that these increments of $\pi $ in $\Phi _g $ should be
interpreted as evidencing a \emph{new} sort of phase, which would be
distinctly related to the topology of $SO(3)$. All these results are indeed
known since a couple of years and have been discussed within a more general
framework by Sj\"{o}qvist \cite{sjoqvist}, who was interested on a variable
degree of entanglement. From a physical point of view, it appears to be far
more interesting to study the nonclassical dependence of the geometric phase
on the degree of entanglement \cite{sjoqvist}, than to investigate the
topology of $SO(3)$ by experimental means. Even so, and because questions
related to the subtleties of the $SO(3)$ topology have recently attracted
much attention, we find it necessary to analyze in detail what features the
newly proposed experiments would have in common with those already performed
in the past using neutrons. In the following section we employ the tools of
quantum optics to analyze some representative experiments on neutron
interferometry.

\section{Neutron interferometry}

A $\pi $-phase that arises as a consequence of a $2\pi $ rotation
applied to spinors is a familiar feature of these eminently
quantum-mechanical objects. The $-1$ factor that the corresponding
wave function acquires as a consequence of such an evolution was
first assumed to be unobservable, but experiments performed by
Werner \emph{et al}. \cite{werner} and by Rauch \emph{et al.}
\cite{rauch} - prompted by \emph{Gedanken} experiments discussed
by Aharonov and Susskind \cite{aharonov} and Bernstein \cite
{bernstein} - brought such a factor into evidence. These
experiments used an unpolarized neutron beam, which was coherently
split into two other beams at the entrance of an interferometer.
The two beams were reflected by means of crystals, so as to bring
them to a point where they recombine and interfere, before being
detected. In order to make the $\pi $-phase appear, one of the two
beams was subjected to the action of a tunable magnetic field,
before recombining it with the other beam. Werner \emph{et al}.
\cite{werner} used two detectors, $D_{1}$ and $D_{2}$, in order to
show the sinusoidal variation of the difference in neutron counts,
$N(D_{1})-N(D_{2})$, as a
function of the magnetic field; that is, as a function of $\omega t$, with $%
t $ fixed and variable $\omega $. The arrangement is essentially a Young's
type interferometer (see Fig. (1)), in which one of the beams is subjected
to a tunable dephasing.

In a conventional Young's interferometer the counting rate is given by an
expression of the form $N(D)\sim 1+k\cos \varphi $, where the constant $k$
encapsulates how the interferometer acts on the incoming beam by splitting
it into two other beams, and $\varphi $ is a phase arising from their path
difference. The essential point that we want to stress here is the
following: a full quantum-mechanical treatment of the Young's interferometer
requires that we invoke a field theoretic description, with the radiation
field being given by a density operator $\rho =\left| \psi \right\rangle
\left\langle \psi \right| $, and the electromagnetic field being described
in terms of annihilation and creation operators for each possible mode. For
single-mode excitation, the Young's interference experiment can be described
\cite{walls} in terms of the boson operator $b^{\dagger }=a_{1}^{\dagger
}\cos \theta +a_{2}^{\dagger }\sin \theta $, where each $a_{i}$ $(i=1,2)$
refers to the respective pinhole of the interferometer depicted in Fig.(1).
The photon state out of which the interference fringes arise is therefore
given by $b^{\dagger }\left| vac\right\rangle =\cos \theta \left|
1,0\right\rangle +\sin \theta \left| 0,1\right\rangle $, with $\left|
n_{1},n_{2}\right\rangle $ meaning a state with $n_{i}$ photons in the
corresponding mode $a_{i}$. The constant $k$ introduced above reads,\ for
the present case \cite{walls}, $k=\sin 2\theta $. For modes of equal
intensity one has $\theta =\pi /4$ and the counting rate becomes $N(D)\sim
1+\cos \varphi $, in accordance with the classical result. We see that a
proper, full quantum-mechanical treatment of Young's experiment requires the
appearance of the \emph{entangled} state $b^{\dagger }\left|
vac\right\rangle =\left( \left| 1,0\right\rangle +\left| 0,1\right\rangle
\right) /\sqrt{2}$. An analogous treatment can be envisaged for neutron
experiments of the sort performed by Werner \emph{et al}. \cite{werner}.
Indeed, as Silverman \cite{silverman} has shown, a Mach-Zehnder
charged-fermion interferometer, in which a confined magnetic field causes an
Aharonov-Bohm phase-shift between the two beams, can be treated field
theoretically like Young's experiment. This is also the case for the
experimental arrangement of Werner \emph{et al}., as one can easily show.
\begin{figure}[tbp]
\begin{center}
\includegraphics[angle=0,scale=.6]{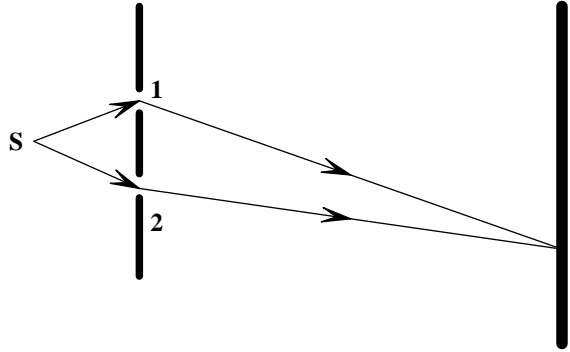}
%\includegraphics[angle=-90,scale=1.0]{multi.eps}
%\end{center}
\end{center}
\caption{Two-slit Young's interferometer.}
\label{f1}
\end{figure}
The interferometer used by Werner \emph{et al}. (Fig. 2) can be described
\cite{yurke} in terms of fermion annihilation operators $c_{a}^{out}(s)$ and
$c_{b}^{out}(s)$, which correspond to the output states going towards
detectors $D_{a}$ \ and $D_{b}$ respectively. These operators satisfy the
usual anticommutation relations and can be written in terms of the input
operators $d_{a}(s)$ and $d_{b}(s)$ in the usual way \cite{yurke,mandel}.
Here, $a$ and $b$ refer to the two spatial modes (i.e., arms) of the
interferometer and $s$ makes explicit the dependence of the operators on the
spin variables. Such an interferometer is effectively the same as a
Mach-Zehnder interferometer like the one shown in Fig.(3). We can describe
it in terms of the parameters characterizing the beam-splitter and the
dephasing elements. We thus obtain, suppressing explicit reference to $s$,
for brevity:

\begin{figure}[ptb]
\begin{center}
\includegraphics[angle=0,scale=.8]{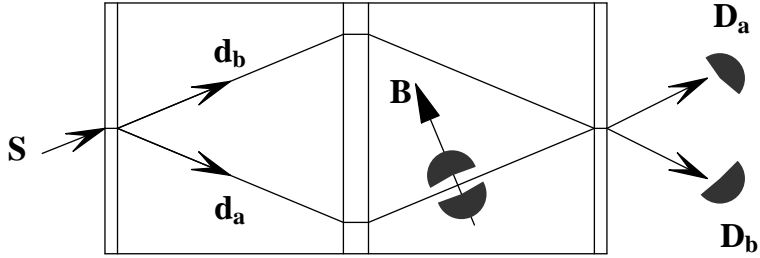}
%\includegraphics[angle=-90,scale=1.0]{multi.eps}
%\end{center}
%
%
\end{center}
\caption{Neutron interferometer. There are three crystal slabs through which
the neutron beams go. The first and third slab play the role of
beam-splitters, whereas the second (middle) slab plays the role of the
mirrors in a Mach-Zehnder interferometer. One beam passes through a tunable,
transverse dc magnetic field $B$ causing spin precession. The other beam
goes through a field-free region. A $2\protect\pi$ precession of the one
beam causes a relative $\protect\pi$-shift between the phases of the two
beams.}
\label{f2}
\end{figure}

\begin{figure}[tbp]
\begin{center}
\includegraphics[angle=0,scale=.8]{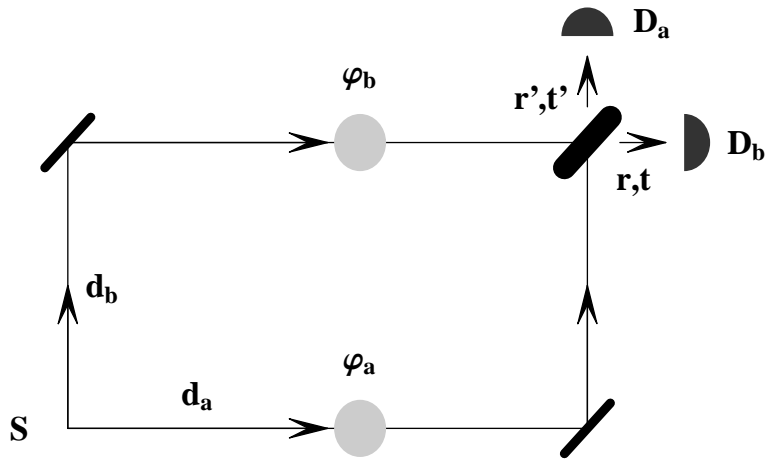}
%\includegraphics[angle=-90,scale=1.0]{multi.eps}
%\end{center}
%
%
\end{center}
\caption{Mach-Zehnder interferometer in which one of the two beam-splitters
has been replaced (or is represented by) a source $S$ of entangled states.
There are two dephasing elements, $\protect\varphi _{a}$ and $\protect%
\varphi _{b}$. One of them can be used to make a $\left( -1\right) ^{n}$
term appear in the coincidence counts of detectors $D_{a}$ and $D_{b}$. The
parameters $r$, $t$, $r^{\prime}$, $t^{\prime}$ (reflection and transmission
amplitudes) characterize the four-ports, lossless beam-splitter.}
\label{f3}
\end{figure}
\begin{equation}
\left(
\begin{array}{c}
c_{a}^{out} \\
c_{b}^{out}
\end{array}
\right) =\left(
\begin{array}{cc}
te^{i\varphi _{a}} & r^{\prime }e^{i\varphi _{b}} \\
re^{i\varphi _{a}} & t^{\prime }e^{i\varphi _{b}}
\end{array}
\right) \left(
\begin{array}{c}
d_{a} \\
d_{b}
\end{array}
\right) ,
\end{equation}
where $\varphi _{i}$ is the phase shift suffered by the beam going along arm
$i=a,b$. The parameters $r$, $t$, $r^{\prime }$, $t^{\prime }$, refer to the
reflection and transmission amplitudes of a four-ports, lossless
beam-splitter.

The counting rates in detectors $D_{a}$ and $D_{b}$ are given by

\bigskip
\begin{equation}
\overline{N(D_{i})}=Tr\left\{ \rho \sum_{s=\uparrow ,\downarrow
}(c_{i}^{out}(s))^{\dagger }c_{i}^{out}(s)\right\} ,\text{ \ }(i=a,b),
\end{equation}
so that, writing them in terms of $d_{a}$ and $d_{b}$, we have

\begin{eqnarray*}
\overline{N(D_{a})} &=&\sum_{s}\left| t\right| ^{2}Tr\left( \rho
d_{a}^{\dagger }(s)d_{a}(s)\right) +\left| r^{\prime }\right| ^{2}Tr\left(
\rho d_{b}^{\dagger }(s)d_{b}(s)\right) +2\left| t^{\ast }r^{\prime
}Tr\left( \rho d_{a}^{\dagger }(s)d_{b}(s)\right) \right| \cos \Phi (s) \\
\overline{N(D_{b})} &=&\sum_{s}\left| r\right| ^{2}Tr\left( \rho
d_{a}^{\dagger }(s)d_{a}(s)\right) +\left| t^{\prime }\right| ^{2}Tr\left(
\rho d_{b}^{\dagger }(s)d_{b}(s)\right) +2\left| r^{\ast }t^{\prime
}Tr\left( \rho d_{a}^{\dagger }(s)d_{b}(s)\right) \right| \cos \Psi (s).
\end{eqnarray*}

Here, $\Phi (s)=\triangle \varphi (s)+\alpha +\beta $, and $\Psi
(s)=\triangle \varphi (s)+\gamma +\beta $, with $\triangle \varphi
(s)=\varphi _{b}(s)-\varphi _{a}(s)$, whereas the other phases are defined
through the following equations: $t^{\ast }r^{\prime }=\left| t^{\ast
}r^{\prime }\right| e^{i\alpha }$, $r^{\ast }t^{\prime }=\left| r^{\ast
}t^{\prime }\right| e^{i\gamma }$, and $Tr\left( \rho d_{a}^{\dagger
}d_{b}\right) =\left| Tr\left( \rho d_{a}^{\dagger }d_{b}\right) \right|
e^{i\beta }$.

The density operator $\rho $ that corresponds to the experimental
arrangement of Werner \emph{et al}. \cite{werner} is build up as a
statistical mixture of pure, antisymmetrized states of the form $\left| \psi
_{s}\right\rangle =(\left| 1_{s},0\right\rangle -\left| 0,1_{s}\right\rangle
)/\sqrt{2}$ , with $s=\uparrow ,\downarrow $ (spin up and down). It is thus
given by $\rho =\sum \frac{1}{2}\left| \psi _{s}\right\rangle \left\langle
\psi _{s}\right| $. On one of the two arms of the interferometer there is a
magnetic field causing the spin precession that gives rise to $\triangle
\varphi (s)$. Depending on whether $s=\uparrow $ or $s=\downarrow $, we have
$\triangle \varphi (s)=\pm \triangle \varphi $, with $\triangle \varphi
=2\pi g_{n}\mu _{N}M\lambda Bl/h^{2}$. We refer to \cite{werner} for the
meaning of the parameters, noting only that $B$ is the variable magnetic
field and $l$ the length over which it acts. With a $\rho $ as given above
we obtain

\bigskip
\begin{eqnarray}
\overline{N(D_{a})} &=&\frac{1}{2}\left( \left| t\right| ^{2}+\left|
r^{\prime }\right| ^{2}+\left| t^{\ast }r^{\prime }\right| (\cos (\alpha
-\triangle \varphi )+\cos (\alpha +\triangle \varphi ))\right) \\
\overline{N(D_{b})} &=&\frac{1}{2}\left( \left| r\right| ^{2}+\left|
t^{\prime }\right| ^{2}+\left| r^{\ast }t^{\prime }\right| (\cos (\gamma
-\triangle \varphi )+\cos (\gamma +\triangle \varphi ))\right)
\end{eqnarray}

The results of Werner \emph{et al}. can be reproduced by appropriately
choosing the beam-splitter parameters, so as to mimic their experimental
arrangement. Besides that, we need also to add a residual phase shift $%
\delta $ that is attributable to various causes, including gravity \cite
{werner,staudenmann}. Taking $\left| t\right| ^{2}+\left| r^{\prime }\right|
^{2}=2C$, $\left| t^{\ast }r^{\prime }\right| =A$, $\alpha =\pi $, and $%
\left| r\right| =\left| t^{\prime }\right| =\sqrt{A}$, $\gamma =0$, we have

\begin{eqnarray}
\overline{N(D_{a})} &=&C-\frac{A}{2}(\cos (\delta -\triangle \varphi )+\cos
(\delta +\triangle \varphi ))  \label{na} \\
\overline{N(D_{b})} &=&\frac{A}{2}\left( 1+\cos (\delta -\triangle \varphi
)\right) +\frac{A}{2}(1+\cos (\delta +\triangle \varphi )),  \label{nb}
\end{eqnarray}
in accordance with the results given by Werner \emph{et al}. (see Eqs.(3)
and (4) of Ref.\cite{werner}). We see therefore that the experiments in
neutron interferometry, showing the sign reversal of a spinor subjected to a
$2\pi $ rotation, can be interpreted as arising from entangled states of the
form $\left| \psi _{s}\right\rangle =(\left| 1_{s},0\right\rangle -\left|
0,1_{s}\right\rangle )/\sqrt{2}$. In what follows, we write states of this
sort as $(\left| \uparrow _{a},0_{b}\right\rangle -\left| 0_{a},\uparrow
_{b}\right\rangle )/\sqrt{2}$, and $(\left| \downarrow
_{a},0_{b}\right\rangle -\left| 0_{a},\downarrow _{b}\right\rangle )/\sqrt{2}
$, in order to stress the analogy with the ones considered by Milman and
Mosseri: ($(\left| H_{a},V_{b}\right\rangle +\left| V_{a},H_{b}\right\rangle
)/\sqrt{2}$).

Experiments like those of Werner \emph{et al}. have been usually interpreted
as putting into evidence the transformation properties of spin-$1/2$
particles under rotations. These rotations are represented by elements of
the $SU(2)$-manifold, the universal covering group of $SO(3).$ The
one-to-one map that can be established is a map between elements of $%
SU(2)=S^{3}$ and matrices $U$ of the form given by Eq.(\ref{2}). These
matrices can be taken as representing a rotation matrix, or else a MES. In
fact, both cases are actually the same - as we pointed out before - because
any MES can be obtained from a given, fixed one - say, $\left|
1,0\right\rangle $ - by applying to it a rotation. There is a \emph{two}%
-to-one map between matrices of the form given by Eq.(\ref{2}) and elements
of $SO(3)$: $U$ and $-U$ both map onto the same $R_{U}=R_{-U}$ in $SO(3)$.

The classical experiments in neutron interferometry have been recently
analyzed in terms of \emph{mixed state phases} by Sj\"{o}qvist \emph{et al }
\cite{sjoqvist2,bhandari}. These authors interpret the sign reversal of a
spinor subjected to a $2\pi $ rotation as a consequence of a phase shift $%
\phi =\pi $, that they define in the following way:

Consider a Mach-Zehnder interferometer. To each of its arms we associate the
state vectors $\left| a\right\rangle $ and $\left| b\right\rangle $,
respectively. In the Hilbert space spanned by these vectors we can represent
any input state, e.g., $\widetilde{\rho }_{in}=\left| a\right\rangle
\left\langle a\right| $, as well as the effect of mirrors, beam splitters,
and relative phase shifts, by $2\times 2$ unitary matrices $\widetilde{U}$.
The output state $\widetilde{\rho }_{out}$ will then be obtained as $%
\widetilde{\rho }_{out}=\widetilde{U}\widetilde{\rho }_{in}\widetilde{U}%
^{\dagger }$, where $\widetilde{U}$ is the product of all those
transformations suffered by the input state when it traverses the
interferometer. If among these transformations there is a relative phase
shift $\chi $, the intensity $I$ of the output beam along $\left|
a\right\rangle $ can be shown \cite{sjoqvist2} to be given by $\left\langle
a\right| \widetilde{\rho }_{out}\left| a\right\rangle $, as $I\sim 1+\cos
\chi $ .

Assume now that the particles carry some internal degrees of freedom, like
spin. Let $\left| k\right\rangle ,$ $k=1,2,\ldots N$, denote the vectors
spanning the internal space. These vectors are assumed to be chosen so that
the associated density operator $\rho _{0}$ is initially diagonal: $\rho
_{0}=\sum_{k}w_{k}\left| k\right\rangle \left\langle k\right| $, $w_{k}$
being the classical probability to find the pure state $\left|
k\right\rangle $ as part of the ensemble. The internal density operator can
change inside the interferometer as described by the unitary transformation $%
\rho _{0}\rightarrow U_{i}\rho _{0}U_{i}^{\dagger }$. The internal state is
assumed to remain unaffected by mirrors, beam splitters, etc., so that the
above defined operators $\widetilde{U}$ are now extended to operators $%
\mathbf{U}=\widetilde{U}\otimes 1_{i}$ that act on the full
Hilbert space with basis $\left\{ \left| j\right\rangle \otimes
\left| k\right\rangle ;j=a,b,\ k=1,2,\ldots N\right\} $.
Correspondingly, an incoming state along the $\left|
a\right\rangle $-arm of the interferometer can be represented by
$\mathbf{\rho }_{in}=\widetilde{\rho }_{in}\otimes \rho
_{0}=\left|
a\right\rangle \left\langle a\right| \otimes \rho _{0}$. We stress that $%
\rho _{0}$ might correspond to a mixed state, in general. In order to define
a generalization of the Pancharatnam phase to the case of mixed states
Sj\"{o}qvist \emph{et al.} \cite{sjoqvist2} introduce the unitary
transformation

\begin{equation}
\mathbf{U}=\left(
\begin{array}{cc}
e^{i\chi } & 0 \\
0 & 0
\end{array}
\right) \otimes 1_{i}+\left(
\begin{array}{cc}
0 & 0 \\
0 & 1
\end{array}
\right) \otimes U_{i},  \label{mixdef}
\end{equation}
which corresponds to applying the phase-shift $\chi $ along path
$\left| a\right\rangle $ and $U_{i}$ along path $\left|
b\right\rangle $. As proved in Ref.\cite{sjoqvist2}, the output
intensity along $\left| a\right\rangle $ is now given by

\begin{equation}
I\sim 1+\nu \cos \left( \chi -\phi \right) ,  \label{mix1}
\end{equation}
where the phase $\phi $ and the ``visibility'' $\nu $ are defined through

\begin{equation}
Tr\left[ U_{i}\rho _{0}\right] =\nu e^{i\phi }.
\end{equation}

Sj\"{o}qvist \emph{et al. }\cite{sjoqvist2} propose to take $\phi =\arg Tr%
\left[ U_{i}\rho _{0}\right] $ as the natural generalization of the
Pancharatnam phase. For a pure state $\rho _{0}=\left| \psi
_{0}\right\rangle \left\langle \psi _{0}\right| $ the phase $\phi $ reduces
to the usual Pancharatnam phase between $U_{i}\left| \psi _{0}\right\rangle $
and $\left| \psi _{0}\right\rangle $. For a mixed state of a qubit, the
corresponding density matrix can be written in terms of the Pauli matrices
as $\rho =\left( 1+r\widehat{\mathbf{r}}\cdot \mathbf{\sigma }\right) /2$,
with $\widehat{\mathbf{r}}$ a unit vector and $r\leq 1$. The pure
eigenstates $\left| \pm ;\widehat{\mathbf{r}}\cdot \mathbf{\sigma }%
\right\rangle $ of $\rho $ satisfy $\rho \left| \pm ;\widehat{\mathbf{r}}%
\cdot \mathbf{\sigma }\right\rangle =\frac{1}{2}\left( 1\pm r\right) \left|
\pm ;\widehat{\mathbf{r}}\cdot \mathbf{\sigma }\right\rangle $ and acquire
Pancharatnam phases $\phi _{\pm }=\mp \Omega /2$ when $\widehat{\mathbf{r}}$
traces out a geodesically closed curve on the Bloch sphere that subtends the
solid angle $\Omega $. The generalized Pancharatnam phase is given in this
case by $\phi =\arg \left( \eta \left( \cos \left( \Omega /2\right) -ir\sin
\left( \Omega /2\right) \right) \right) =-\arctan \left( r\tan \left( \Omega
/2\right) \right) $; $\eta $ being the common visibility of the two pure
states: $\nu _{+}=\nu _{-}\equiv \eta $. For a maximally mixed state we have
$r=0$, $\eta =\left| \cos \left( \Omega /2\right) \right| $ and $\phi =\arg
\left( \cos \left( \Omega /2\right) \right) $. In this case, the intensity $%
I $ can be written as

\begin{equation}
I\sim 1+\left| \cos \left( \frac{\Omega }{2}\right) \right| \cos \left[ \chi
-\arg \cos \left( \frac{\Omega }{2}\right) \right] =1+\cos \chi \cos \left(
\frac{\Omega }{2}\right) .  \label{mix2}
\end{equation}

This formula shows that when $\Omega =2\pi $ there is a sign change that can
be traced back to the phase shift $\phi =\arg \cos \pi =\pi $ . This way,
the $4\pi $ symmetry of spinors, as it was tested using unpolarized
neutrons, can be related to the generalized Pancharatnam phase. Eq.(\ref
{mix2}) can be compared with Eq.(\ref{nb}), which can be rewritten as $I\sim
1+\cos \delta \cos \triangle \varphi $. Our $\delta $ corresponds to the
relative phase-shift $\chi $ between the beams in each arm of the
interferometer and which is attributable - as already said - to different
causes, including gravity, while our $\triangle \varphi $ corresponds to $%
\Omega /2$, the phase introduced by acting on the internal degrees of
freedom, i.e., on the spin, by means of a magnetic field.

At first sight, the above interpretation of the neutron experiments seems to
be rather unrelated to the one given in terms of entangled states. In order
to see how these two interpretations relate to one another, we resort to an
alternative description \cite{sjoqvist2} of the output intensity, Eq.(\ref
{mix2}), given in terms of pure state interference profiles $I_{k}$. These
profiles arise as follows. Consider the pure input state $\left|
k\right\rangle $. In accordance with Eq.(\ref{mixdef}), if this state goes
along the $\left| a\right\rangle $-arm of the interferometer, it suffers the
change $\left| k\right\rangle \rightarrow \exp (i\chi )\left| k\right\rangle
$, whereas if it goes along the $\left| b\right\rangle $-arm it suffers the
change $\left| k\right\rangle \rightarrow U_{i}\left| k\right\rangle $. The
output intensity associated with the input state $\left| k\right\rangle $ is
then given by

\begin{equation}
I_{k}\sim \left| e^{i\chi }\left| k\right\rangle +U_{i}\left| k\right\rangle
\right| ^{2}\sim 1+\left| \left\langle k\right| U_{i}\left| k\right\rangle
\right| \cos \left( \chi -\arg \left\langle k\right| U_{i}\left|
k\right\rangle \right) .  \label{mix3}
\end{equation}

Now, as we have already seen in the case of Young's interferometer, the
output intensity, as given by $I\sim 1+k\cos \varphi $, can also be derived
within the framework of a full quantum description. Such a description
requires, however, that we resort to entangled states. The same holds true
in the case of Eq.(\ref{mix3}) and with it, also in the case of a weighted
mixture of pure states, as given by $\rho _{0}=\sum_{k}w_{k}\left|
k\right\rangle \left\langle k\right| $. This mixed state $\rho _{0}$ gives
rise to an output intensity $I=\sum_{k}w_{k}I_{k}$. It is then easy to see
that this output intensity profile is the same as the one given by Eq.(\ref
{mix2}):

\begin{eqnarray}
I &=&\sum_{k}w_{k}I_{k}\sim 1+\sum_{k}w_{k}\left| \left\langle
k|U_{i}k\right\rangle \right| \cos \left( \chi -\arg \left\langle
k|U_{i}k\right\rangle \right) \\
&=&1+\sum_{k}w_{k}\nu _{k}\cos \left( \chi -\phi _{k}\right) =1+\nu \cos
\left( \chi -\phi \right) .
\end{eqnarray}

Here, $\nu _{k}=\left| \left\langle k\right| U_{i}\left| k\right\rangle
\right| $, $\phi _{k}=\arg \left\langle k\right| U_{i}\left| k\right\rangle $%
, $\phi =\arg \left( \sum_{k}w_{k}\nu _{k}\exp \left( i\phi _{k}\right)
\right) =\arg \left[ Tr\left( U_{i}\rho _{0}\right) \right] $, $\nu =\left|
\sum_{k}w_{k}\nu _{k}\exp \left( i\phi _{k}\right) \right| =\left| Tr\left(
U_{i}\rho _{0}\right) \right| $. This way we see that the seeked connection
between the two interpretations occurs at the pure-state level, where a full
quantum mechanical approach requires a treatment in terms of entangled
states.

Next, let us calculate the number of coincident counts, $\overline{%
N(D_{a})N(D_{b})}$, for the density matrix given above. We obtain the
following result:

\begin{eqnarray*}
\overline{N(D_{a})N(D_{b})} &=&Tr\left\{ \rho
\sum_{s}(c_{a}^{out}(s))^{\dagger }c_{a}^{out}(s)\sum_{s^{\prime
}}(c_{b}^{out}(s^{\prime }))^{\dagger }c_{b}^{out}(s^{\prime })\right\} \\
&=&\frac{1}{2}\left\{ \left| rt\right| ^{2}+\left| r^{\prime }t^{\prime
}\right| ^{2}+2Re\left[ rr^{\prime }(tt^{\prime })^{\ast }\right] \right\} \\
&&-\frac{1}{2}\left\{ e^{i\triangle \varphi }\left[ r^{\ast }t^{\prime
}\left| t\right| ^{2}+t^{\ast }r^{\prime }\left| t^{\prime }\right| ^{2}%
\right] +e^{-i\triangle \varphi }\left[ t^{\prime \ast }r\left| r^{\prime
}\right| ^{2}+tr^{\prime \ast }\left| r\right| ^{2}\right] \right\} .
\end{eqnarray*}

The parameters of a lossless beam-splitter have to satisfy the equations
\cite{mandel} $\left| r\right| ^{2}+\left| t\right| ^{2}=1$, and $r^{\ast
}t+t^{\prime \ast }r^{\prime }=0$, from which it follows that $t^{\prime
}=t^{\ast }$ and $r^{\prime }=-r^{\ast }$. Replacing these values in the
above, general expression, we see that $\overline{N(D_{a})N(D_{b})}=0$, as
it must, because our mixed state $\rho $ has been constructed out of
one-particle states. Now, our aim is to compare a coincidence count in a
neutron experiment with the one obtained with an arrangement like the one
proposed by Milman and Mosseri. To this end, let us consider the case of a
pure state $\rho =\left| \psi \right\rangle \left\langle \psi \right| $,
with $\left| \psi \right\rangle $ given by a \emph{two}-fermion, singlet
state $\left| \psi \right\rangle =\left( \left| \uparrow _{a}\downarrow
_{b}\right\rangle -\left| \downarrow _{a}\uparrow _{b}\right\rangle \right) /%
\sqrt{2}$.

Furthermore, let us consider an experimental arrangement like the one shown
in Fig.(3), i.e., a Mach-Zehnder interferometer in which one of the two
beam-splitters has been replaced by a source $S$ of the aforementioned
entangled states $\left| \psi \right\rangle =\left( \left| \uparrow
_{a}\downarrow _{b}\right\rangle -\left| \downarrow _{a}\uparrow
_{b}\right\rangle \right) /\sqrt{2}$. We obtain in this case (taking $%
r^{\prime }=r$ and $t^{\prime }=t^{\ast }$, i.e., parameters corresponding
to a symmetrical beam-splitter):

\begin{equation}
\overline{N(D_{a})N(D_{b})}=\left| t\right| ^{4}+\left| r\right|
^{4}+2\left| rt\right| ^{2}\cos \left( \triangle \varphi _{+}-\triangle
\varphi _{-}\right) ,
\end{equation}
where $\triangle \varphi _{+}=\varphi _{b}(\uparrow )-\varphi _{a}(\uparrow
) $, and $\triangle \varphi _{-}=\varphi _{b}(\downarrow )-\varphi
_{a}(\downarrow )$, so that $\triangle \varphi _{+}-\triangle \varphi
_{-}=\left( \varphi _{b}(\uparrow )-\varphi _{b}(\downarrow )\right) -\left(
\varphi _{a}(\uparrow )-\varphi _{a}(\downarrow )\right) \equiv \triangle
\varphi _{b}-\triangle \varphi _{a}$. We have assumed that, in general, both
dephasings may depend on the spin orientation. If we want to consider a case
like the one corresponding to Fig.(2) it suffices to take the dephasing in
arm $b$ as being independent on the spin orientation, $\varphi _{b}(\uparrow
)=\varphi _{b}(\downarrow )$, whereas $\varphi _{a}(\uparrow )-\varphi
_{a}(\downarrow )=\triangle \varphi _{a}$. As for the beam-splitter
parameters, let us take again the values corresponding to the experiments of
Werner \emph{et al}. that we chose before: $\left| t\right| ^{2}+\left|
r^{\prime }\right| ^{2}=2C$, $\left| t^{\ast }r^{\prime }\right| =A$, and $%
\left| r\right| =\left| t^{\prime }\right| =\sqrt{A}$. We obtain the
following expression for the number of coincident counts:

\begin{equation}
\overline{N(D_{a})N(D_{b})}=2A^{2}\left\{ 1+\cos \left( \triangle
\varphi_{a} \right) \right\}.
\end{equation}

If we instead consider that both dephasings, $\varphi _{a}$ and $\varphi
_{b} $, depend on the spin orientation, we can arrange the experiment so
that the dephasing in one arm, $a$ say, becomes an integer multiple of $\pi$
so as to have $\triangle \varphi _{a}=n\pi $. In such a case we obtain,
considering again $\left| r\right| =\left| t^{\prime }\right| =\sqrt{A}$,

\begin{eqnarray}
\overline{N(D_{a})N(D_{b})} &=&2A^{2}\left\{ 1+(-1)^{n}\cos \left( \triangle
\varphi _{b}\right) \right\} \\
&=&4A^{2}\cos ^{2}\left( \triangle \varphi _{b}/2\right) \text{, \ \ for }n%
\text{ even} \\
&=&4A^{2}\sin ^{2}\left( \triangle \varphi _{b}/2\right) \text{, \ \ for }n%
\text{ odd}.
\end{eqnarray}

Thus, we see that the coincidence rate will depend on $n$ being even or odd,
as in the proposal of Milman and Mosseri. The present analysis should serve
to make clear some points that in the later proposal could have caused some
confusion. In the proposal of Milman and Mosseri there are three dephasing
elements (properly oriented wave plates) put on three arms (modes $a$, $b$
and $c$), respectively. The rate of coincident counts in modes $a$ and $b$
is predicted to be given by $P=(1/2)\left| \left( -1\right) ^{n}-\cos \phi
\right| $, where $\phi $ comes from a tunable dephasing element put on one
arm, $b$, of a Mach-Zehnder interferometer. The term $\left( -1\right) ^{n}$
comes from the dephasing introduced in the opposite arm, $c$, of the
Mach-Zehnder interferometer and it should depend on the nature of the
trajectory \cite{milman} in $SO(3)$. There is a third dephasing element, put
on arm $a$, which lies outside the Mach-Zehnder interferometer. Now, it is
the relative phase between the two arms of a Mach-Zehnder interferometer,
the one upon which counting rates usually depend. One could therefore expect
that a third arm is needed in order to make an extra term $(-1)^{n}$ appear.
The arrangement depicted in Fig.(3) shows that the appearance of such a term
depends instead on the possibility to drive two phases independently from
one another. A third mode seems to be unnecessary, even as a reference mode,
so that the arrangement shown in Fig.(3) could be used in the case of
experiments with entangled photons, as well.

\section{Conclusion}

We have analyzed in detail different interferometric arrangements in terms
of entangled states. In particular, we focussed our attention on the
Mach-Zehnder interferometer as a basic tool, with the help of which some
subtleties of the $SO(3)$ topology can be experimentally displayed.
Experiments performed in the past with neutrons are here re-analyzed in
terms of entangled states. We compared our analysis with another one which
has been recently put forward, based on the geometric phase for mixed
states. We showed that in both pure and mixed cases a full
quantum-mechanical treatment requires the introduction of entangled states.
This is so because the basic physical phenomenon behind an interferometric
experiment, which is the interference pattern arising out of two states,
does require the introduction of entangled states when field quantization
has been invoked. This remains true for mixed states, because they are build
up as a statistically weighted mixture of pure states.

We have also addressed the question as to what new features would come out
by performing experiments like those proposed by Milman and Mosseri or by
LiMing \emph{et al}. A central point of their proposals is the one-to-one
mapping between the elements of the set $\Omega _{MES}$ and the elements of $%
SO(3)$. This bijection is achieved by the identification $\left( \alpha
,\beta \right) \sim \left( -\alpha ,-\beta \right) $ used in the definition
of the set $\Omega _{MES}$. However, one thing is to mathematically define
such a set and quite another thing is to experimentally implement it. To
this end, one should be able to give a prescription as to how is it possible
- at least in principle - to realize the afore said identification in a
laboratory. As we stressed it before, the identification that is
experimentally realized routinely is the one between $(\alpha ,\beta )$ and
the whole ray $e^{i\psi }(\alpha ,\beta )$.

The experiments proposed by Milman and Mosseri consider subjecting one of
the two beams that constitute a MES to a variable dephasing. The other beam
serves as a reference one, with respect to which the phase-shift of the
first beam is measured. This shift may be engineered so as to be dependent
on an even or an odd multiple of $\pi $, by splitting it into two beams with
the help of a Mach-Zehnder interferometer. In so far, the physics upon which
the effect relies is just the same as in a classical neutron experiment: a
relative $\pi $-phase difference arising out of a rotated two-level system.
It is immaterial that such a two-level system be realized in practice as a
spin-$1/2$ particle (a neutron) or as a polarized photon. Entanglement is in
both cases behind the observed phenomenon. We have considered a neutron
experiment designed to measure coincidence counts when one introduces
independent dephasings in the two arms of a Mach-Zehnder interferometer.
These coincidence counts would depend on the $\pi $-phase gained by a spinor
under a $2\pi $ rotation. Such an experiment would require to feed the
Mach-Zehnder interferometer with a singlet state of two fermions. To our
knowledge, these experiments have not been performed. In the context of the
classical neutron experiments, which were designed to reveal the $\pi $%
-phase, it would have been superfluous to perform such experiments with
entangled, two-fermion states. Of course, their realizability with the
technology that was available three decades ago is quite another question.

Present day technology offers the opportunity to develop experiments that
complement those performed in the past with neutrons. The newly proposed
experiments would make use of entangled states of two-level systems -
polarized photons - which are not straightforwardly realized with spin-$1/2$
particles, like neutrons or electrons. Arrangements of the sort proposed by
Milman and Mosseri \ - or by LiMing \emph{et al}. - widen markedly the
versatility for realizing different paths along the $SU(2)$-manifold, when
compared with the classical neutron experiments. However, to consider the
predicted effects as a manifestation of a \emph{new} topological phase seems
to be, in our view, unjustified. In any case, the simple arrangement we
discussed in the present work, which is illustrated in Fig.(3), should serve
to make clear which essential features could be brought into evidence by
experiments of this sort.

\section{Appendix}

Here we remind some known facts \cite{gilmore,sattinger} related
to topological properties of the groups $SO(3)$ and $SU(2)$.
Whereas these properties are well known, in the view of some
authors \cite{milman} they could lead to misinterpretations that
can be found even in textbooks. It seems thus justified to include
here a very short review of some results concerning Lie groups, in
order to make sure that our statements be unambiguous.

Let us first discuss in some detail the double connectedness of
$SO(3)$. The Lie groups $SU(2)$ and $SO(3)$ are locally
isomorphic, although globally they differ from each other. Whereas
$SU(2)$ is simply connected, $SO(3)$ is not. Both groups share one
and the same algebra. Among the groups sharing a given algebra,
only one is simply connected. This group is called the
\emph{universal covering group}. All groups sharing a Lie algebra
can be obtained from their universal covering group. To this end,
one needs to determine the discrete invariant subgroups $D$ of the
simply connected group
$SG$. The elements $d_{i}\in D$ satisfy $gd_{i}g^{-1}=d_{i}$ for all $g$ in $%
SG$. After having determined $D$, one can construct the factor group $SG/D$.
The factor group has the same Lie algebra as $SG$. When $D$ has more than
one element, $SG/D$ is multiply connected. In our case $SG$ is $SU(2)$, and
its discrete invariant subgroup $D$ is $Z_{2}=\{I,-I\}$, where $I$
represents the identity. Hence, $SO(3)$ is obtained from $SU(2)$ as the
factor group $SU(2)/Z_{2}$.

\begin{figure}[tbp]
\begin{center}
\includegraphics[angle=0,scale=.6]{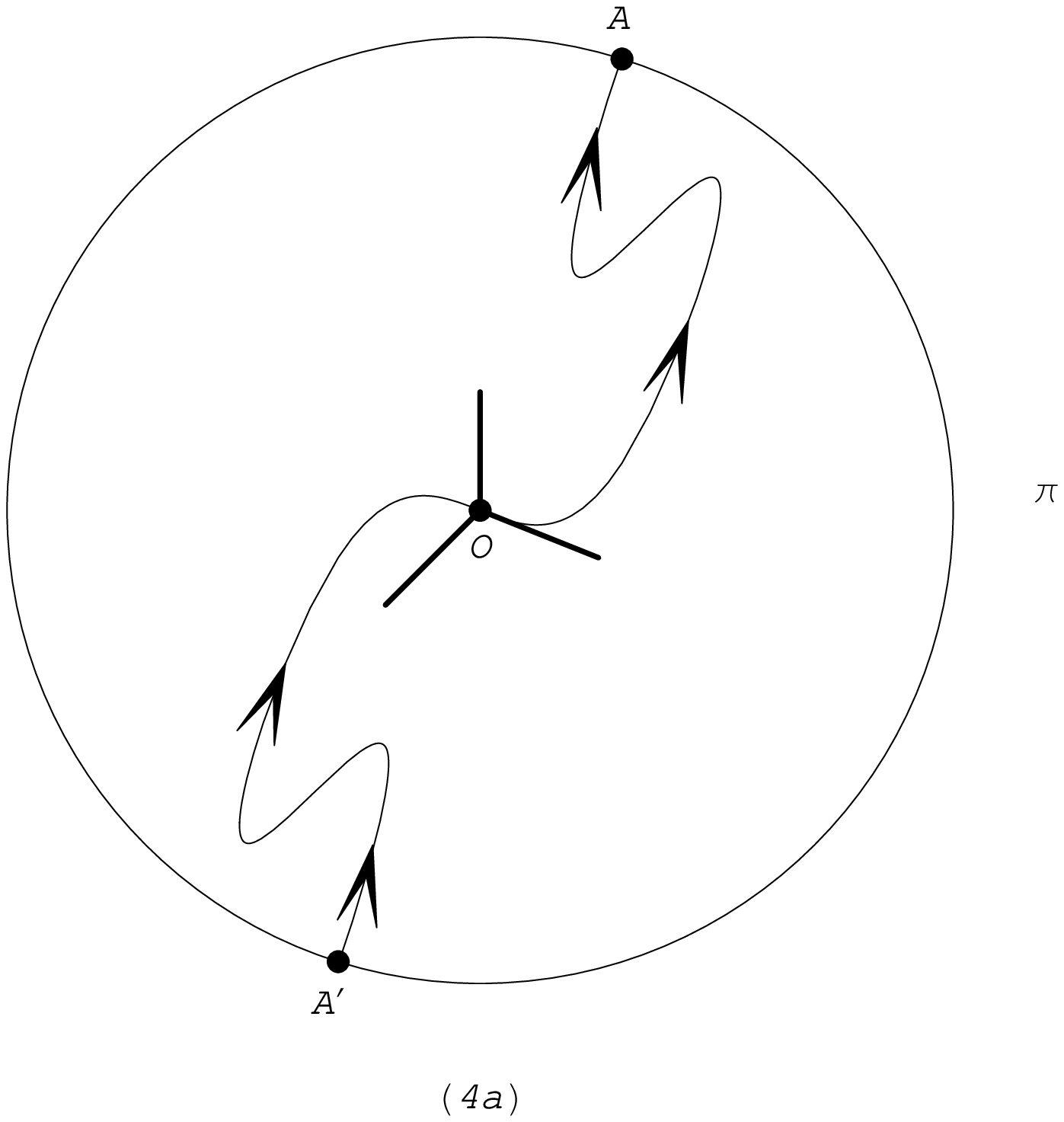}
%\includegraphics[angle=-90,scale=1.0]{multi.eps}
%\end{center}
\end{center}
\caption{The parameter space of $SO(3)$ can be represented as a
solid ball
of radius $\protect\pi$. Each point $P$ of this ball at a distance $\protect%
\theta $ from the origin represents a counterclockwise rotation about axis ${%
OP}$ by an angle $\protect\theta $. Antipodal points represent the same $%
SO(3)$ rotation. The paths shown in (a) and (b) are homotopically
equivalent. They start and end at the origin (the identity) $O$,
but cannot be continuously reduced to the identity transformation.
$A$ and $A^{\prime }$ correspond to a single group operation in
$SO(3)$. In $SU(2)$, however, rotations about
$\widehat{\mathbf{z}}$ and $-\widehat{\mathbf{z}}$ can be
distinguished from one another.} \label{f1a}
\end{figure}

\begin{figure}[ptb]
\begin{center}
\includegraphics[angle=0,scale=.6]{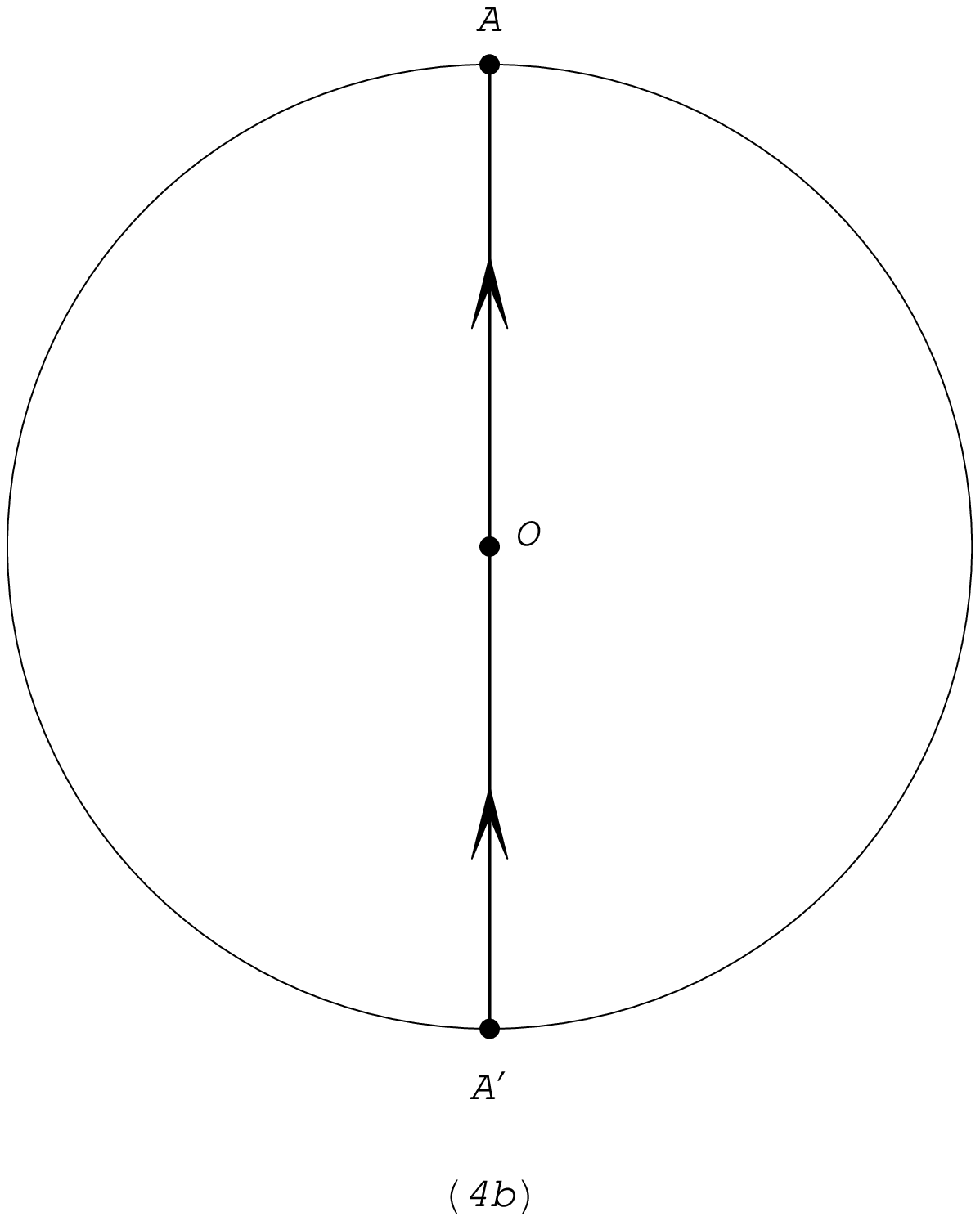}
%\includegraphics[angle=-90,scale=1.0]{multi.eps}
%\end{center}
\end{center}
\caption{Same caption as in Fig.(\ref{f1a}).} \label{f1b}
\end{figure}

It is well known that the elements of a Lie group can be obtained from the
elements of its algebra by exponentiation. To any curve in the parameter
space of the algebra it corresponds a curve in the parameter space of the
group. Conventionally, one takes the origin of the parameter space of the
group as representing the identity element. The group $SO(3)$ is a
three-parameter group. It can be visualized by picturing its elements as
points of a solid ball of radius $\pi $. Each point $P$ of this ball at a
distance $\theta $ from the origin represents a \emph{counterclockwise}
rotation about the axis $\overrightarrow{OP}$ by an angle $\theta $. Hence,
antipodal points on the surface of the ball represent the same rotation. For
example, in Fig.(\ref{f1a}) points $A$ and $A^{\prime }$ represent the same
rotation. The path shown in Fig.(\ref{f1a}) corresponds to the application
of successive rotations. It starts with the identity transformation $O$
until it reaches the rotation represented by $A=A^{\prime }$, from where it
goes back to $O$. Although $A$ and $A^{\prime }$ represent the same
rotation, we should stress that to the jump from $A$ to $A^{\prime }$ it
corresponds a change in the rotation axis. The path shown in Fig.(\ref{f1b})
is homotopically the same as the one shown in Fig.(\ref{f1a}); i.e., one of
these paths can be continuously deformed into the other. The path shown in
Fig.(\ref{f1b}) corresponds to a series of rotations sharing the common axis
$\widehat{\mathbf{z}}$.

\begin{figure}[ptb]
\begin{center}
\includegraphics[angle=0,scale=.6]{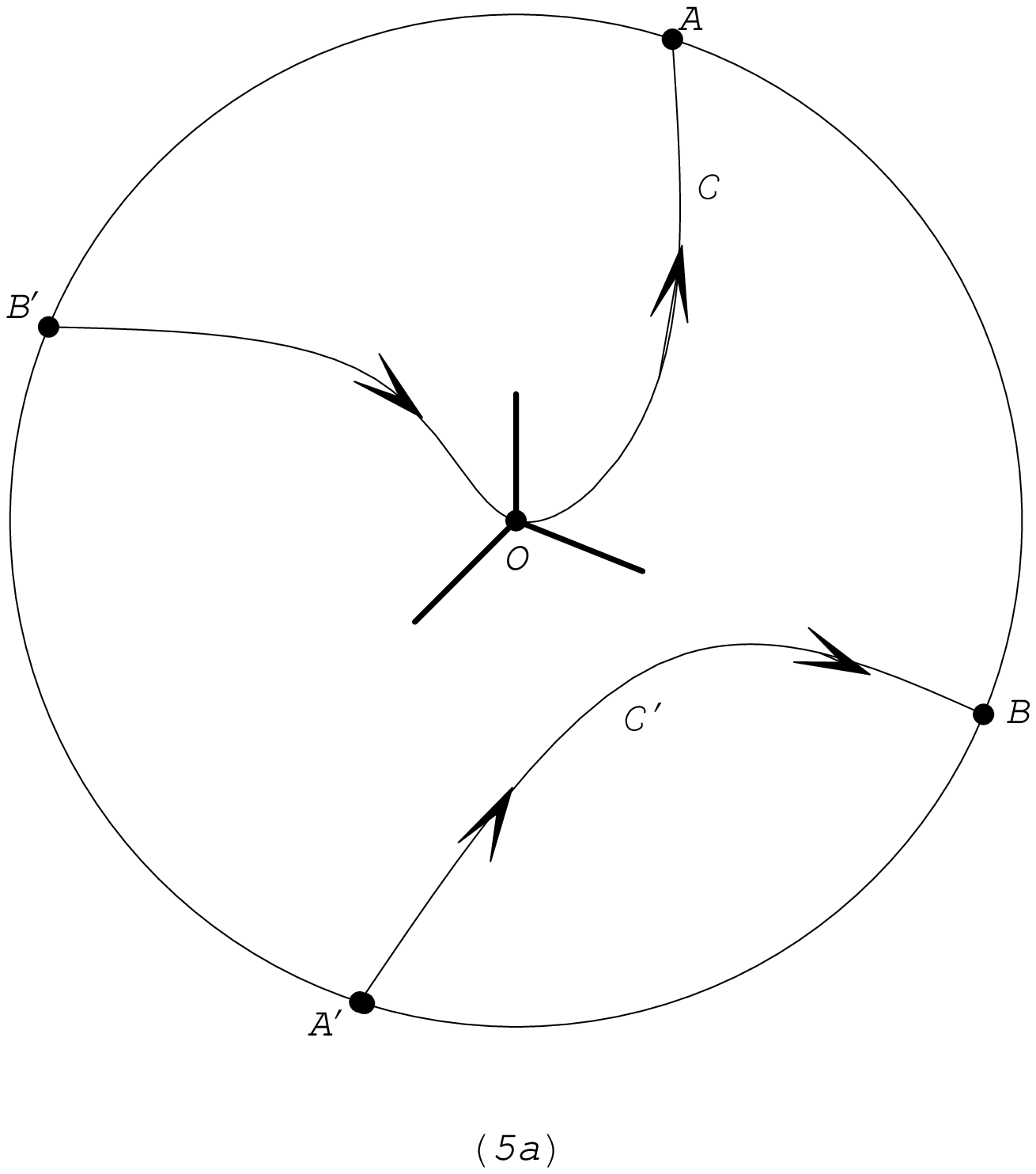}
%\includegraphics[angle=-90,scale=1.0]{multi.eps}
%\end{center}
\end{center}
\caption{The path shown in (a) starts and ends at $O$. It consists
of a sequence of rotations that go as $O\rightarrow A=
A^{\prime}\rightarrow B=B^{\prime} \rightarrow O$. Such a path can
be continuously deformed to the identity. Part (b) of the figure
shows an intermediate step towards a deformation to the identity.
This path belongs to a homotopy class which is different from the
one to which the path shown in Fig.(\ref{f1a}) belongs. There are
two such homotopy classes in $SO(3)$, corresponding to the double
connectedness of the group.} \label{f2a}
\end{figure}

\begin{figure}[ptb]
\begin{center}
\includegraphics[angle=0,scale=.6]{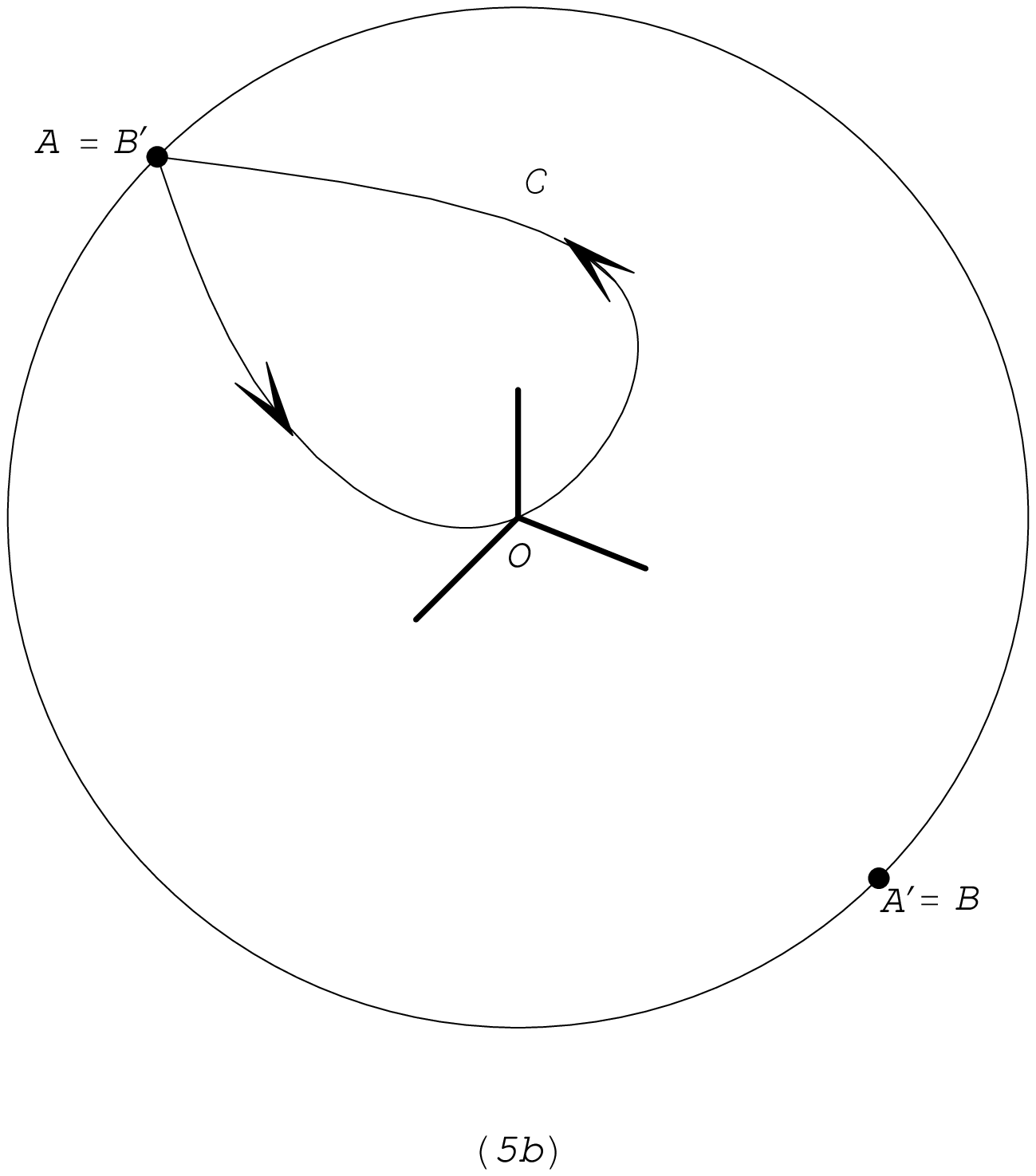}
%\includegraphics[angle=-90,scale=1.0]{multi.eps}
%\end{center}
\end{center}
\caption{Same caption as in Fig.(\ref{f2a}).} \label{f2b}
\end{figure}

The fact that $SO(3)$ is doubly connected can be evidenced by considering a
path like the one shown in Fig.(\ref{f2a}). It represents a series of
rotations going from $O$ to $A=A^{\prime }$, from there to $B=B^{\prime }$,
from where one goes back to $O$. Such a path is homotopically the same as
the paths shown in Figs.(\ref{f1a}) and (\ref{f1b}), when these paths have
been traversed \emph{twice}. In other words, the path shown in Fig.(\ref{f1b}%
) traversed twice can be continuously deformed to the path shown in Fig.(\ref
{f2a}). This last path, in turn, can be continuously deformed to the
identity. Fig.(\ref{f2b}) shows an intermediate step of such a
transformation. Any curve like $\mathcal{C}$ is homotopically equivalent to
the identity. On the contrary, a curve like the one shown in Fig.(\ref{f1a})
is not. The double connectedness of $SO(3)$ is related to the fact that
there exist paths like this last one, which cannot be deformed into a single
point at $O$. On the other hand, the parameter space of $SU(2)$ consists of
a solid ball of radius $2\pi $, for which all points on the surface
represent the same transformation: $-I$. Any closed path starting and ending
at $O$ can be continuously deformed to a single point at the origin. $SU(2)$
is thus simply connected.

\begin{figure}[ptb]
\begin{center}
\includegraphics[angle=0,scale=.6]{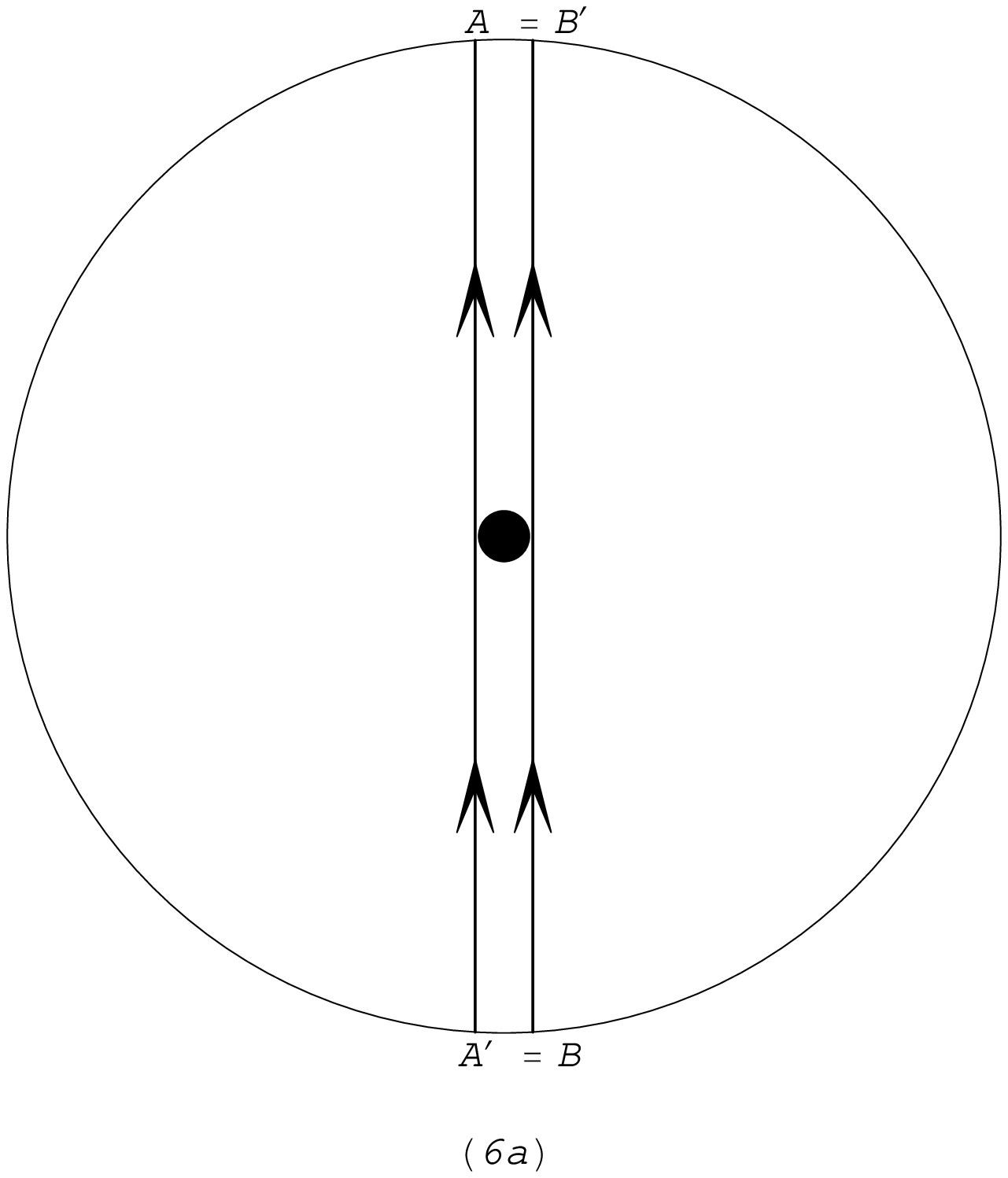}
%\includegraphics[angle=-90,scale=1.0]{multi.eps}
%\end{center}
\end{center}
\caption{The path depicted in (a) represents a sequence of rotations about $%
\widehat{\mathbf{z}}$. For clarity, this axis has been split into
two lines. The path shown corresponds to the one of
Fig.(\ref{f1a}) when it is traversed twice. The path in this
figure is then homotopically the same as the one of
Fig.(\ref{f2a}) and can be reduced to the identity. The path shown
in (b) can be deformed to the identity, as well. However, (a)
corresponds to a sequence of rotations about
$\widehat{\mathbf{z}}$, whereas in (b) half of the path
corresponds to rotations about $-\widehat{\mathbf{z}} $.}
\label{f3a}
\end{figure}

\begin{figure}[tbp]
\begin{center}
\includegraphics[angle=0,scale=.6]{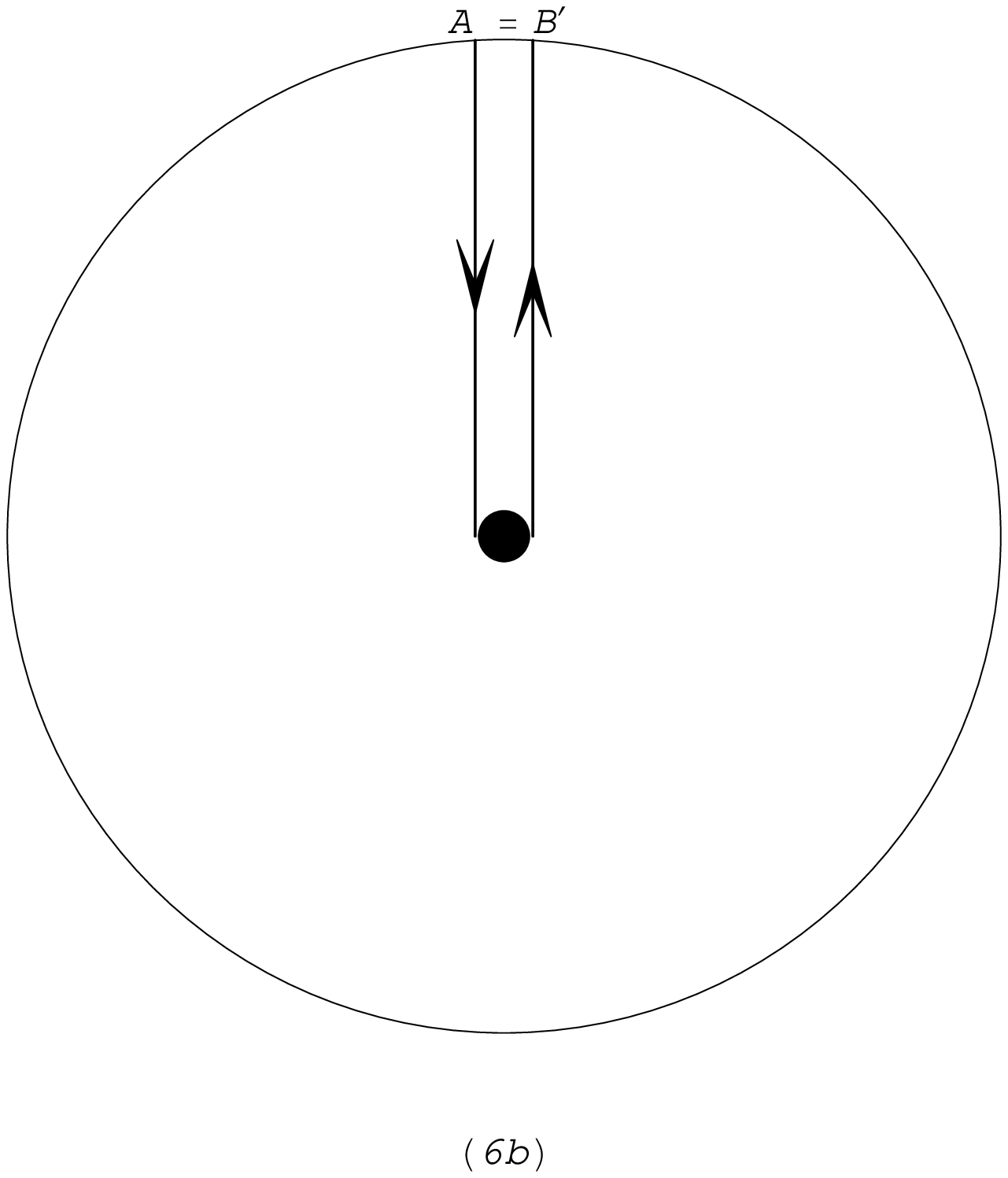}
%\includegraphics[angle=-90,scale=1.0]{multi.eps}
%\end{center}
\end{center}
\caption{Same caption as in Fig.(\ref{f3a}).} \label{f3b}
\end{figure}

The double connectedness of $SO(3)$ comes about in situations like the one
illustrated in Fig.(\ref{f3a}). It represents a series of transformations
that start and end at the identity: $O\rightarrow A=A^{\prime }\rightarrow
O\rightarrow B^{\prime }=B\rightarrow O$. For clarity, we have displaced one
curve from the other, although what is meant is a series of rotations about
the same axis $+\widehat{\mathbf{z}}$ by the same angle $\pi $. The total
angle covered along this path is therefore $4\pi $ and, as we have seen
before, such a path is homotopically identical to the identity. The path
shown in Fig.(\ref{f3b}): $O\rightarrow B^{\prime }=A\rightarrow O$ is
homotopically identical to the identity, as well. However, in this case the
first half of the path corresponds to a $\pi $-rotation about $+\widehat{%
\mathbf{z}}$ and the second one to a $\pi $-rotation about $-\widehat{%
\mathbf{z}}$. The two paths of Figs.(\ref{f3a}) and (\ref{f3b}) pertain to
the same homotopy class. On the contrary, paths like the ones shown in Figs.(%
\ref{f1a}) and (\ref{f1b}) correspond to the second homotopy class. They
cannot be deformed to the identity $I$, but are associated with the other
element of $Z_{2}$, which is $-I$.

We are now ready to address the question as to how could we design
an experiment that brings into evidence the double connectedness
of $SO(3)$. According to what we saw, the answer reads as follows:
If we want to disclose the subtleties of $SO(3)$ which are related
to its double connectedness, what we need is a physical effect
that clearly distinguishes the transformations $I$ and $-I$ from
one another. We need an object for which a $2\pi $ rotation would
bring about a minus sign. Such an object is, of course, a
spin-$1/2$ particle. Hence, it is precisely in order to disclose
the subtleties of $SO(3)$ that we need to work with its universal
covering group, which is $SU(2)$. This was the goal of the neutron
experiments \cite{werner,rauch} performed three decades ago. An
experiment designed to disclose the subtleties of $SO(3)$ should
preferably map elements of $SU(2)$ into physical operations, not
into states. In our case, the operation corresponds to an element
of $SU(2)$ and it is applied to a spinor. It is essential that
this spinor be part of an entangled state, in order to obtain an
interference pattern from which a relative phase may be read off.

\newpage

\end{document}